\documentclass[12pt]{article}
\newcommand{\spzg}{\hspace{15mm}}
\newcommand{\be}{\begin{equation}}
\newcommand{\ee}{\end{equation}}
\newcommand{\bea}{\begin{eqnarray}}
\newcommand{\eea}{\end{eqnarray}}
\newcommand{\ton}[1]{\left( #1 \right)}
\newcommand{\qua}[1]{\left[ #1 \right]}

\newcommand{\goto}{\rightarrow}

\def\acq{a_q^\dagger}
\def\aq{a_q}
\def\ra{\rangle}
\def\la{\langle}

\def\eps{\epsilon}

\def\disp{\displaystyle}

\def\Z{{\bf Z}}

\newcommand{\re}{{\rm e}}
\begin{document}
\oddsidemargin 5mm
\setcounter{page}{0}
\renewcommand{\thefootnote}{\fnsymbol{footnote}}
\newpage     
\setcounter{page}{0}
\begin{titlepage}
\vspace*{1.cm}
\begin{center}
{\Large {\bf Area versus length distribution \\
for closed random walks}}\\  
\vspace{1.5cm}
{\bf Filippo Colomo}\ \footnote{e-mail: colomo@fi.infn.it}\\
\vspace{0.8cm}
{\em I.N.F.N., Sezione di Firenze \\
and Dipartimento di Fisica, Universit\`a di Firenze, \\ 
Via G. Sansone 1, 50019 Sesto Fiorentino (FI), Italy}\\  
\end{center}
\vspace{2.cm}

\begin{abstract}
Using a connection between the $q$-oscillator algebra and
the coefficients  of the high temperature expansion of the 
frustrated Gaussian spin model, we derive  an exact formula 
for the number of closed random walks  of given length and 
area, on a hypercubic lattice, in the limit of infinite 
number of dimensions. The formula is investigated in detail,
and asymptotic behaviours are evaluated. The area distribution 
in the limit of long loops is computed. As a  byproduct, we obtain
also an infinite set of new, nontrivial identities.
\end{abstract}
\vfill
{\hfill PACS: 05.40.Fb, 05.50.+q}
\vspace{5mm}
\end{titlepage}


\section{Introduction}

Let us consider on an hypercubic lattice of dimension $D$
all closed paths of length $2k$, starting from a given vertex.
An interesting and still unanswered question concerns  the counting of 
such paths according to the area they enclose. 
Besides being a challenging combinatorial problem on its own right,
the question  has some relevance also from a physical point of view:
considering for instance the hopping of a charged particle around 
a two dimensional lattice in a uniform magnetic field 
\cite{harper,zak,hofstadter}, its energy spectrum  displays an 
intricate hierarchic structure \cite{abanov}, whose properties are 
closely related to the area versus length distribution of closed
paths \cite{lipan}. Such distribution finds applications also  
in polymer physics \cite{brereton,duplantier}, and  in explaining 
anomalous magnetoconductance \cite{minkov}.
A different but somehow related problem is the one of counting
random walks according to the number of distinct sites they visit 
\cite{stanley,nieuwenhuizen}: relatively 
compact loops are expected to enclose a small area and to visit
relatively few sites, and viceversa. This counting  problem is relevant
when investigating random walks on lattices with static traps.

Coming back to the initial question of counting loops according
to their enclosed area,  in the case of  a two-dimensional lattice 
the answer was given only in the limit
of asymptotically large $k$ in \cite{levy}, and the first subleading 
correction was provided in \cite{bellissard}. In higher dimension, the
problem is in general even more difficult.
Here we want to address the question in the limit of infinite dimensionality 
of the lattice, a  very peculiar situation,  in which
however some simplifications occur, and an exact answer can be given.
The latter is to be interpreted as  a ``mean-field'' approximation
to the exact counting on a lattice of finite dimensionality $D$.
Such a formulation of the problem was first considered 
by Parisi et al. \cite{parisi1,parisi2}. They were investigating 
spin models with frustration but without any  quenched disorder, in 
order to test the conjecture that such deterministic models could 
behave at low temperature as some suitably chosen spin-glass model  
with quenched disorder.
They considered the frustrated Spherical and XY spin models in the limit 
of large
dimensionality $D$ of the lattice, where the saddle point approximation 
becomes exact. In their analysis of these models, they showed how  the high
temperature expansion (i.e. loop expansion) can be nicely rewritten
by using the $q$-oscillator algebra  \cite{kulish}, 
where $q$ measures the frustration per 
plaquette. 
Similar $q$-deformed algebraic relations have also appeared in the 
Hofstadter problem of quantum particles hopping on a two-dimensional
lattice in a uniform magnetic field \cite{wiegmann}. This problem is 
closely related  to the frustrated spin models, which can be considered 
as simplified models of hopping in the large $D$ and classical limits.
The use of somehow related non-commutative geometry techniques
appeared as a crucial ingredient also in Ref.  \cite{bellissard}.

The problem we address and solve in this paper may be  stated as follows:
given an infinite dimensional hypercubic lattice, what is the number
$G_{k,l}$ of loops  of length $2k$, starting from a given vertex,  
enclosing  a minimal area of exactly $l$ plaquettes\footnote{When 
facing this problem of area versus length distribution of closed random
walks, the area is most often considered in its algebraic sense, i.e.
as oriented; in our
case, however, an average over orientations is intrinsecally performed
in the approach (see Section 2), and we are therefore referring
to the absolute value of the area.}?

The paper is organized as follows. In next Section we summarize some 
results of \cite{parisi1,parisi2}
which are then used in Section 3 to obtain an exact expression
for the counting numbers $G_{k,l}$. In Section 4 a generating 
function for these counting numbers is presented. In Section 5 a probability
distribution is naturally associated to the $G_{k,l}$'s, 
and a systematic procedure is built to compute its moments of arbitrary
order. The asymptotic behaviour of this probability distribution
for large $k$ is evaluated in Section 6, while
Section 7 is devoted to the presentation of some new, 
nontrivial identities. The last Section contains our conclusions, 
while some technical details are relegated into the Appendix.

\section{Frustrated Lattice and $q$-oscillator Algebra}

In Refs. \cite{parisi1,parisi2}, Parisi et al. considered the 
frustrated Gaussian, Spherical and XY models on a $D$-dimensional hypercubic 
lattice, in the large $D$ limit. Among other things, they unveiled a 
remarkable connection between the coefficients of the high
temperature expansion (i.e., loop expansion) and the $q$-oscillator 
algebra  \cite{kulish}, where $q$ measures the frustration per 
plaquette  and varies continuously on the real interval $[-1,+1]$, 
between the fully-frustrated case ($q=-1$, fermionic algebra) and the
ferromagnetic case ($q=1$, bosonic algebra).
For our purposes, the relevant part of the Hamiltonians describing the 
models considered in Refs. \cite{parisi1,parisi2} is:
\be
H=-{1\over\sqrt{2D}}\ \sum_{\langle j k \rangle}
\phi^\dagger_j U_{jk}\phi_k \  + \  {\rm h.c.} \ .
\ee
The complex field $\phi_j\in {\bf C} $ is defined on the sites 
(labelled by $j$) of a $D$-dimensional hyper-cubic lattice; 
frustration is induced on the lattice through the 
nearest-neighbour couplings $U_{jk}$, which are  complex numbers of 
modulus one and satisfy the relation $U_{jk}=U^*_{kj}$; they are the 
link variables of a background Abelian lattice gauge field, chosen in such
a way as to 
produce  a static and constant magnetic field, suitably oriented 
to give the same magnetic flux $\pm B$ for
any plaquette of the lattice (the product of the four $U$'s
around the plaquette being  $e^{\pm i B}$). Such a magnetic field having the 
same projection over  all the axes, up to the sign, in Refs. 
\cite{parisi1,parisi2} these signs were chosen randomly  in order to 
avoid the selection of a preferred direction. 
The usual unfrustrated ferromagnetic spin interaction is obtained 
for $B=0$. Non-vanishing values of $B$ induce a frustration
around each plaquette, which is maximal for $B=\pi$,
the fully-frustrated case. 

In the framework of the high temperature
expansion, the free energy of such models is expressed 
as a sum over the contribution of loops of increasing length $2k$:
\be 
\beta F=\sum_{k=0}^\infty\ {\beta^{2k}\over 2k}\ G_k \ .
\label{htexp}
\ee
Each loop encloses a number of plaquettes;
in the case of the models considered in \cite{parisi1,parisi2}, for each 
loop the magnetic field yields a weight,  proportional to $\exp(iBA)$, 
where $A$ is the sum of plaquettes
with signs depending on the orientations. The total contribution of 
all loops of length $2k$ is given by $G_k$.
Due to the average over orientations and loops, the quantity 
$G_k$ is a polynomial in the variable
$q=\cos B$, the coefficient $G_{k,l}$ of $q^l$ being given by 
the number of loops
of length $2k$ and area $l$ \footnote{The area of a  loop is 
defined as the minimal area of a surface of lattice plaquettes 
having that loop  as  boundary.}.
The infinite dimensionality of the lattice ensures that for any given
loop, no two plaquettes of the subtended minimal surface will lie on the 
same plane; this in turn accounts for such a simple averaging
over orientations, performed indipendently for each plane.
The order of polynomial $G_k(q)$ is
given by ${k(k-1)/2}$, the maximal area encloseable by a loop of 
length $2k$. 
The coefficients $G_{k,l}$ can also be interpreted as the number
of  Feynman diagrams with $2k$ external points, which are joined pairwise 
by lines (propagators) intersecting $l$ times. 
These diagrams also occur in the topological (large $N$) expansion of
Matrix Models \cite{brezin}, where  the planar limit corresponds to no 
intersections, i.e. to the $q=0$ case.
Equivalently, in simple graphical terms, they just can 
be seen as the number of way of connecting pairwise $2k$ points on a 
circle with  $k$  segments intersecting
exactly $l$ times. In the following we shall refer to 
this last picture\footnote{This formulation of the  
problem was considered e.g. by Linus Pauling \cite{pauling} when trying 
to simplify the calculation of matrix elements involved in Slater's
treatment of the electronic structure of molecules.}.
In Ref. \cite{parisi1}, the enumeration of such diagrams was investigated.
In particular, a recursion
relation was found for the coefficients of the polynomial $G_k(q)$
-- a sort of Wick theorem -- which
can be nicely  expressed in terms of  the algebra of the $q$-oscillators  
$a_q, a_q^\dagger$:
\be
\aq\acq- q \acq\aq=1 \ .
\label{qalg}
\ee 
These operators \cite{kulish} act on the Hilbert space spanned by the vectors: 
$|m\ra$, $m=0,\ 1,\ ... $, as follows,
\bea
\aq |m\ra\ =&\sqrt{[m]_q}\quad |m-1\ra \ ,\spzg&
\aq |0\ra= 0 \ ,\nonumber\\
\acq |m\ra\ =& \sqrt{[m+1]_q}\ |m+1\ra \ ,\spzg&
[m]_q=\frac{\  1-q^m}{\  1-q\ } \ .
\eea
Using the recursion relation, the weighted multiplicities of the diagrams 
of eq. (\ref{htexp}) were neatly written as an expectation value over 
the ground state of 
the $q$-oscillators  \cite{parisi1,parisi2}:
\be
G_k(q) \ =\ \la 0|(\acq+\aq)^{2k}|0\ra \ .
\label{qvev}
\ee

The authors of Refs. \cite{parisi1,parisi2} did not exploit
the consequences of this result, preferring to turn themselves to numerical 
investigation of the models they were interested in. As a consequence,
till now only the two limiting cases $G_k(0)$ and $G_k(1)$ were 
explicitly known; when $q=1$ it is only a matter of counting the way
of connecting $2k$ points on a circle, with no restrictions, and this 
is simply the number of pairings of  $2k$ objects: $(2k-1)!!$ .
When $q=0$ we are in fact evaluating  the planar limit
of the zero-dimensional  $2k$-point Green function of a Matrix Model
in the limit of vanishing interaction \cite{brezin}. In simple graphical 
terms this correspond to the number of way of joining pairwise $2k$ points on
a circle with no intersection. In other words, this is just 
one of the many possible definition of Catalan numbers (see, e.g.
\cite{stanleybook}), given as:
\be
G_k(0)=\frac{(2k)!}{k!(k+1)!}\ .
\ee
But the number $G_{k,l}$ of way of connecting pairwise $2k$ points on a circle,
with exactly $l$ intersections (i.e. the coefficient of $q^l$ in $G_k(q)$)
is not so easily accessible.
In Refs. \cite{parisi1,parisi2}, they were found by direct enumeration 
of the graphs on a computer. In \cite{capcol},
a generating function  for coefficients $G_{k,l}$ was obtained,proposed,
but then the explicit evaluation of such  coefficients relied upon
heavy symbolic manipulations.

\section{Exact Solution of the Enumeration Problem\label{solution}}

We shall now present a simple, easy to evaluate, formula for a 
generic coefficient $G_{k,l}$.
To this purpose let us first introduce the $x_q$ 
coordinate representation, $x_q=\acq+\aq$, $x_q|x\ra = x|x\ra$, which is 
given by the so-called continuous $q$-Hermite 
polynomials \cite{gasper,szego}.
These are defined by:
\be
H_n(x) =  \la x | n \ra\ {\cal C}_n\ ,
\qquad {\cal C}_n = \left(\left[n\right]_q!\right)^{1/2}\ {\cal C}_0\ ,
\ee
where the normalization constant ${\cal C}_0$ is fixed by
$H_0(x) =1 $ and the $q$-factorial is 
\be
\left[n\right]_q! =[n]_q \ [n-1]_q \dots [1]_q\ , \qquad
[1]_q=[0]_q=1 \ .
\ee
These polynomials satisfy, of course, a three-term recursion relation
in the index $n$: 
\be
x H_n(x) = H_{n+1}(x) +[n]_q\ H_{n-1}(x)\ , \qquad n\ge 1 \ .
\ee
$x$ ranges  over the interval 
$x\in\left[-2/\sqrt{1-q},2/\sqrt{1-q}\right]$, and a convenient 
convenient parametrisation is
\be
x=\frac{2}{\sqrt{1-q}}\cos\theta \  ,\spzg \theta\in [0,\pi ].
\label{variasm}
\ee
More properties of these $q$-Hermite polynomials can be found
in  Ref.  \cite{gasper}, where they are defined as 
${\cal H}_n(\cos\theta)=(1-q)^{n/2} H_n (x)$.
The most important property for us is the orthogonalizing measure
$\nu_q(x)$  \cite{maassen,gasper,szego}:
\be
\int_{-2/\sqrt{1-q}}^{2/\sqrt{1-q}}\ \nu_q(x)\ dx\ 
H_n(x)\ H_m(x) \ = \  \delta_{n,m}\ [n]_q! \ ,
\ee
\bea
\nu_q(x) &=&\frac{\sqrt{1-q}}{2\pi}\  q^{-1/8}\  
\Theta_1\left({\theta\over\pi},\tau\right)\  \nonumber\\
&=&\frac{\sqrt{1-q}}{\pi}\ \sum_{n=0}^{\infty}\ (-1)^n q^{n(n+1)/2}
\ \sin[(2n+1)\theta]\ , 
\label{qmeasure}
\eea
where $\Theta_1(z,\tau)$ is the first Jacobi theta function \cite{tata}:
\be
\Theta_1\left(z,\tau\right)=\sum_{n\in\Z}\ 
\re^{i\pi\tau(n+\frac{1}{2})^2+ 2\pi i (n+\frac{1}{2})(z-\frac{1}{2})}\ ,
\spzg q=\re^{2\pi i \tau}\ .
\label{theta}
\ee
It is worth emphasizing that, interestingly enough, measure (\ref{qmeasure}) 
interpolates continuously between Wigner semi-circle law for Gaussian 
Matrix Models, and Gaussian distribution, as $q$ varies from $0$ to $1^-$;
this reminds us of the behaviour  of  the  density
distribution of eigenvalues for Gaussian Ensemble of Random Matrices,
as the dimension of the matrices, $N$, varies from $\infty$ to $1$.

Polynomial $G_k(q)$ may now be rewritten as
\be
G_k(q)= \int_{-2/\sqrt{1-q}}^{2/\sqrt{1-q}} \ \nu_q(x) \ x^{2k}\ dx \ . 
\label{gkq}
\ee
Indeed, using explicit form (\ref{qmeasure}) for  the integration measure,
performing the integral and playing a little bit
with indices, we may write
\be
G_k(q)=\ton{\frac{1}{1-q}}^k \sum_{l=0}^{k}(-1)^l\ton{{{2k+1}\atop{k-l}}}
\ {{2l+1}\over{2k+1}} \ q^{l(l+1)/2}\ .
\label{gklsum}
\ee

We now only need to perform explicitly the division of the polynomial
of degree $k(k+1)/2$ defined by the sum in the previous formula.
To this purpose, let us state the following 

{\bf Theorem}\footnote{We are 
unfortunately unable to quote any reference. We  are of course convinced
this theorem has been enounced long time ago, indeed it is just a corollary
of Ruffini's rule. The only proof we have been
able to give, absolutely inelegant and
unworthy of being presented here, is by direct verification.}:
{\em Let $P(q)=\sum_{l=0}^{N} p_l q^l$ be an integer coefficient
polynomial of degree 
$N$ in $q$, exactly divisible by $(1-q)^k$. Then $A(q)=P(q)/(1-q)^k$ 
is an integer coefficient  polynomial
of degree $N-k$ whose coefficients are simply expressed
in terms of the $p_l$'s as follows:
\be
A(q)=\sum_{l=0}^{N-k} a_l q^l\ ,\spzg 
a_l=\sum_{i=0}^{l}\ton{{k+l-1-i}\atop{k-1}}p_i \ .
\label{theorem}
\ee}

\medskip\noindent We therefore readily get a simple closed 
expression for $G_{k,l}$:
\be
G_{k,l}=\sum_{i=0}^{i_{max}} (-1)^i \ton{{k+l-1-i(i+1)/2}\atop{k-1}}
\ton{{2k+1}\atop{k-i}}\frac{2i+1}{2k+1}\ , \spzg l\leq \frac{k(k-1)}{2}\ ,
\label{gkl}
\ee
where $i_{max}= \qua{(-1+\sqrt{1+8l})/2}_{i.p.}$ is the largest integer 
$i$ satisfying $i(i+1)/2 \leq l$.
Because of their definition as counting numbers, all integers $G_{k,l}$ 
should be positive. This fact is indeed not apparent from 
(\ref{gkl}), and  we are not able to present a rigorous proof of this 
statement,
nevertheless we are convinced of its validity. Sensibleness
arguments rely upon the intrinsic positivity of (\ref{qvev})
and its derivatives with respect to $q$, for any value of $q$ in the interval
$[0,1)$.

We have computed explicitly with Mathematica \cite{math}, through 
formula  (\ref{gkl}), which is very efficient, all
values $G_{k,l}$ for $k$ ranging from $1$ to $9$, and verified that they 
indeed match the results of Ref. \cite{parisi2} which were found by 
direct enumeration of the graphs on a computer. 

\section{Generating Function}

Let us define a generating function for the combinatorial numbers $G_{k,l}$:
\be
R(z,q) \equiv \sum_{k=0}^{\infty} G_k(q) z^k
= \sum_{k,l=0}^{\infty} G_{k,l} \ q^l z^k \ .
\ee
The running of index $l$ may or may not be restricted to the range
$0\leq l \leq k(k-1)/2$, without loss of generality:
coefficients $G_{k,l}$ automatically vanish for $l>k(k-1)/2$.
From expression (\ref{gkq}), we may readily
write a generating functions for coefficients $G_k(q)$:
\be
R(z,q)= \int_{-2/\sqrt{1-q}}^{2/\sqrt{1-q}} \ \nu_q(x) \ \frac{1}{1-z x^2}
\ dx \ . 
\label{generating}
\ee
Inserting  for the measure its explicit expression (\ref{qmeasure}),
and performing the integral, we obtain:
\be
R(z,q)=\sqrt{\frac{1-q}{z}}\ \sum_{n=0}^{\infty}\ (-1)^n \ q^{n(n+1)/2} 
\ \qua{K\ton{\frac{z}{1-q}}}^{2n+1}\ ,
\label{generating2}
\ee
where we introduced 
\be
K(\alpha)=\frac{1-\sqrt{1-4\alpha}}{2\sqrt{\alpha}}\ .
\ee
For $q=0$ function (\ref{generating2}) reduces to the standard
generating function of Catalan numbers. The singularities of $R(z,q)$
in the complex $z$-plane are completely determined by those  of the 
simpler function $K(z/(1-q))$. Actually the series is very well 
convergent for $|q|<1$ as any Jacobi theta function. Moreover, for 
$|q|=1$, it is a geometric series, which is still convergent because
$|K(\alpha)|<1$ for $|z|<z_c\equiv (1-q)/4$.

Generating function (\ref{generating2}), as shown in Ref. 
\cite{capcol} in a different context, can be interpreted physically
as the internal energy at temperature $T=1/\sqrt{z}$
of a Gaussian spin model defined on the sites of the infinite-dimensional 
frustrated lattice introduced in Ref. \cite{parisi1,parisi2}.

\section{Moments of the Distribution\label{moments}}

The interpretation of  coefficients $G_{k,l}$ naturally suggests
to introduce, for each positive integer value of index $k$, the following 
normalized (discrete) probability distribution:
\be
P_k(l)= \frac{G_{k,l}}{\sum_{l=0}^{k(k-1)/2} G_{k,l}}=\frac{G_{k,l}}{G_k(1)}
=\frac{G_{k,l}}{(2k-1)!!}\ ,
\spzg \left\{
\begin{array}{l}
k=0, 1, \dots \\ 
l=0,1,\dots,k(k-1)/2\ ,
\end{array}
\right.
\label{distrib}
\ee
whose interpretation is obvious: given an arbitrary diagram with $2k$ legs,
the probability for it to have exactly $l$ crossings is simply given
by $P_k(l)$; equivalently, $P_k(l)$ measures the probability,
for a randomly chosen closed path of length $2k$ on our infinite
dimensional hypercubic lattice, to have area equal 
to $l$.
The moments of the previous probability distribution are defined
as
\be
M^{(k)}_s \equiv \la l^s \ra_k =
\sum_{l=0}^{k(k-1)/2} l^s P_k(l)\ ,
\ee
and can be computed from the following generating function
\be
{\tilde P}_{k}(q)\equiv \la q^l \ra_k = \frac{G_{k}(q)}{G_{k}(1)}\ .
\ee
Indeed, 
\be
M^{(k)}_s = \qua{\ton{q\frac{\partial\ }{\partial q}}^s{\tilde P}_{k}(q)}_{q=1}
=\qua{\ton{\frac{\partial\ }{\partial\log q}}^s{\tilde P}_{k}(q)}_{q=1}\ .
\label{momenta}
\ee

The moments of the deviations with respect to the mean $M^{(k)}_1$ 
(central moments) are given by a similar expression:
\be
m^{(k)}_s\equiv \la (l-M_1^{(k)})^s\ra_k =
\qua{\ton{\frac{\partial\ }{\partial\log q}}^s 
{\tilde P}_{k}(q) \ q^{- M_1^{(k)}}
}_{q=1}\ .
\ee
The computation of such moments, besides being interesting on its own
right, is (at least for the first two) a necessary step towards
the evaluation of the asymptotic form
of probability distribution (\ref{distrib}) in the limit $k\goto\infty$.
We therefore need a systematic procedure to calculate $G_k(q)$, together
with its derivatives, at $q=1$. Indeed, from expression (\ref{gkq}), 
we essentially need to evaluate the derivatives at $q=1$ of the integration 
measure (\ref{qmeasure}). Unfortunately such a limit, due to the presence
of the Jacobi Theta Function, is highly singular: for fixed values of
$x$, $\nu_q(x)$ develops an essential singularity at $q=1$.
A prescription should be given, in our situation the natural one being
to chose to approach $q=1$ from real, smaller than $1$,  values of $q$;
in a more convenient, equivalent notation, we want to evaluate the
behaviour of $\nu_q(x)$ in the limit $\eps \goto 0^+$, $\eps$ real, defined 
by $q\equiv\re^{-\eps}$.
With such a choice, as shown in the Appendix, we can write
any given moment $M^{(k)}_s$ as a simple combination of trivial
Gaussian moments:
\be
M_s^{(k)}=\frac{(-1)^s}{(2k-1)!!}\frac{1}{\sqrt{2\pi}}
\int_{-\infty}^{+\infty}
\ dx\ x^{2k}\ c_s(x)\ \re^{-\frac{x^2}{2}}\ ,
\label{explicitmoment1}
\ee
where $c_s(x)$ is an even polynomial of degree $4s$ in $x$. The only 
practical difficulty is the explicit computation
of the $s$'th coefficient $c_s(x)$ of the Taylor expansion around 
$\eps=0^+$, of the function  
\be
\sqrt{1-\re^{-\eps}} \re^{\frac{\eps}{8}}
\frac{1}{\sqrt{\eps}}\re^{-\frac{1}{2}\qua{\frac{4}{\eps}\arcsin^2
\ton{\frac{x}{2}\sqrt{1-\re^{-\eps}}}-x^2}}\ .
\label{explicitmoment2}
\ee
The expansion can be done by hand  up to the first few orders in $\eps$, 
but soon becomes intractable. It is however a trivial task if we use some 
symbolic manipulation program such as Mathematica\cite{math}.
We shall give here   explicit results for the first few moments:
\bea
M_0^{(k)}&=& 1\nonumber\\
M_1^{(k)}&=& k(k-1)/6\nonumber\\
M_2^{(k)}&=& k(k-1)(5k^2-k+12)/180\\
M_3^{(k)}&=& k(k-1)(35 k^4+14k^3+235k^2-188k+24)/7560\nonumber\\
M_4^{(k)}&=& k(k-1)(175 k^6+315 k^5+2341 k^4-1959 k^3+2056 k^2 
-5664 k-2160)/226800
\nonumber
\eea
Analogously, the  evaluation of  the central moments $m_s^{(k)}$, is
reduced to the  explicit computation
of the $s$'th coefficient of the Taylor expansion around 
$\eps=0^+$, of the function  
\be
\sqrt{1-\re^{-\eps}} \ \re^{\frac{\eps}{8}}\ 
\frac{1}{\sqrt{\eps}}\ \re^{-\frac{1}{2}\qua{\frac{4}{\eps}\arcsin^2
\ton{\frac{x}{2}\sqrt{1-\re^{-\eps}}}-x^2}}\ \re^{\eps k(k-1)/6}\ ,
\label{explicitmoment3}
\ee
with the following results: 
\bea
m_0^{(k)}&=& 1\nonumber\\
m_1^{(k)}&=& 0\nonumber\\
m_2^{(k)}&=& (k+3)k(k-1)/45\\
m_3^{(k)}&=& (2k+3)(2k+1)k(k-1)/945\nonumber\\
m_4^{(k)}&=& k(k-1)(7 k^4+37 k^3 +7 k^2-108 k -45)/4725
\nonumber
\eea

By direct inspection
of the formulae, it is possible to extract the large $k$ behaviour of  
generic moments. Since $c_s(x)$ is a polynomial of degree $4s$ in $x$, 
$M_{s}^{(k)}\sim (2k+4s-1)!!/(2k-1)!! =O(k^{2s})$ as 
$k\goto\infty$\footnote{More precisely, it is easy to show that
$M_{s}^{(k)}= (k^{2}/6)^s[1+O(1/k)]$ for large $k$.}.
Following a similar line of thought, and taking into account
some cancellations which may occur more or less severely, 
according to the parity of $s$, it is also possible to guess 
that $\disp{m_s^{(k)}=O(k^{[\frac{3}{2}s]_{i.p.}}})$ 
for  asymptotically large $k$.

Let  us further notice  that, on the  infinite dimensional hypercubic 
lattice, from $M_1^{(k)}\sim k^2/6$, the area
of random loops increases on average as  the square of their
length. This should be contrasted  to the two-dimensional lattice
\cite{bellissard}, where the average area increases of course as
the square of the average linear extension of the loop, that is as  the
first power of the length itself.

We shall conclude this Section by considering coefficients $G_{k,l}$
for fixed values of $l$. We then have for each integer value $l$ an
infinite succession of integer numbers.
From direct inspection of formula (\ref{gkl}),  for asymptotically 
large values of $k$ we can write:
\be
G_{k,l}=G_{k,0} \ \frac{k^l}{l!}\qua{1+O\ton{\frac{1}{k}}}\ ,
\spzg l\ {\rm fixed.}
\label{fixedl}
\ee

\section{Asymptotic Behaviour of Distribution $P_k(l)$}

Let us turn back to  probability distribution $P_k(l)$, eq. (\ref{distrib}).
We want to investigate its asymptotic behaviour as $k\goto\infty$.
The standard procedure is to rescale the discrete index $l$
to a continuous variable $t$ using the average and the standard deviation
as computed in Section \ref{moments} for generic $k$:
\be
l\ \goto\ \mu_k+\sigma_k t\ ,\spzg 
{{\mu_k\equiv M_1^{(k)}=\frac{k(k-1)}{6}}\ ;
\atop{\sigma_k^2\equiv m_2^{(k)}=\frac{(k+3)k(k-1)}{45}\ .}}
\label{rescaling}
\ee
It is now possible to define the asymptotic  probability
distribution (\ref{distrib}) as:
\be
P(t) =\lim_{k\goto\infty}  \ \sigma_k \ P_k(\mu_k +\sigma_k t) \ .
\label{asymprob}
\ee

Unfortunately the coefficients $G_{k,l}$, are defined
in terms of alternate sums of very large numbers (about $\sqrt{2l}$ terms,
each one of
order $O(k^k)$, summing up to a much smaller number \footnote{Just as an
example,  let us quote that $G_{100,1000}$ is expressed through formula
(\ref{gkl}) as an alternate sum of 45 numbers of order up to $10^{200}$, 
being itself of order $10^{179}$ i.e. 21 orders  of magnitude smaller.}),
so that the evaluation of the asymptotic distribution $P(t)$ is not
completely straightforward. Indeed, we are interested
into the behaviour of  the $G_{k,l}$'s for large $k$ and large $l$,
$l\sim\ k^2/6$, and the simple estimate of  eq. (\ref{fixedl}) is now
completely useless.

In principle the asymptotic behaviour of coefficients $G_k(q)$ for 
large value of $k$ is encoded in the behaviour of  generating 
function $R(z,q)$, eq. (\ref{generating2}), in the vicinity of its closest 
(with respect to the origin) singularity in the variable $z$.
The latter is situated at $z=z_c\equiv(1-q)/4$. But since we are
in fact interested into the behaviour of $G_{k,l}$ where $l$
is also asymptotically large, but in a controlled way, see 
(\ref{rescaling}), we must inspect the generating function in the vicinity
of its closest (again, with respect to the origin) singularity
in variable $q$, that is near $q=1$. But in this way 
$z_c$ is collapsing towards the origin, conspiring to the construction
of an essential singularity in $z=z_c=0$ as $q\goto 1^- $.
To tackle this difficulty we shall not investigate the generating 
function $R(z,q)$, but we shall rather resort to the Laplace 
transform method, applied directly to probability distribution
$P_k(l)$; let us first introduce the Laplace transform
\be
\phi_k(s)=\sum_{l=0}^{k(k-1)/2} \ P_k(l)\  
\re^{-s\ton{\frac{l-\mu_k}{\sigma_k}}}
\ee
of  probability distribution $P_k(l)$, suitably shifted and rescaled 
according to eq. (\ref{rescaling}).
In the large $k$ limit, $\phi_k(s)$ is expected to converge pointwise to the 
Laplace transform of distribution $P(t)$:
 \be
\int_{-\infty}^{+\infty} dt \ P(t) \ \re^{-s t}\ .
\label{laplace2}
\ee
Since we may write
\be
\phi_k(s)= \re^{s\frac{\mu_k}{\sigma_k}} \ \la \re^{-s\frac{l}{\sigma_k}}\ra
= \re^{s\frac{\mu_k}{\sigma_k}} 
\ \frac{ G_k(\re^{-\frac{s}{\sigma_k}})}{G_k(1)}\ ,
\ee
and $\sigma_k\sim k^{3/2}$ for large $k$, the asymptotic 
behaviour of $P_k(l)$  is ruled  by the behaviour of
$G_k(q)$ in the neighbourhood of unity. Indeed, after identification of 
$\eps$ with $s/\sigma_k$, we can make use of the machinery developed
in Section \ref{moments} and Appendix A.
We start by using the fact that, up to terms vanishing exponentially 
for small $\eps$, ($q=\re^{-\eps}$)
\be
G_k(q)=\int_{-\infty}^{+\infty} \ dx \ x^{2k} {\cal F}(x,\eps)\ ,
\ee
with  ${\cal F}(x,\eps)$ given by equation (\ref{effedixeps}).
For $\eps=0$ ($q=1$), the evaluation of the integral reduces to  a 
straightforward 
application of the saddle-point method: we have to sum up the contribution
of the neighbourhoods of two stationary points situated in 
$x=x_{\pm}=\pm\sqrt{2k}$, respectively. We readily get
\be
G_k(1) =  \sqrt{2} \ (2k)^{k} \ \re^{-k} \ \qua{1+O\ton{\frac{1}{k}}}\ ,
\ee
which is indeed the leading asymptotics for $(2k-1)!!$, as it should.
Now, for $\eps$ small but positive (we are therefore specializing to 
positive real values of $s$), for each of the two stationary
point $x=x_{\pm}$, we just  shift the integration variable
$x$ to $x_{\pm}+y$ and expand the integrand, including ${\cal F}(x,\eps)$,
around $k=\infty$.
Considering e.g. the contribution coming from 
the positive stationary point $x_+$, and taking into account only
terms up to $O(\log k)$, we get
\be
I_+ \sim \frac{1}{\sqrt{2\pi}}\int dy \ 
\re^{\qua{k\log 2k -k -y^2 -\frac{s}{2}\sqrt{5k}-\sqrt{10} sy-2s^2}} \ ,
\spzg (s>0).
\ee
Of course the situation in which $s<0$ has to be computed, too,
but the result turns out to be independent of the sign of $s$.
Integrating, and adding  the analogous contribution  coming from the 
region in the neighbourhood of $x_-$, we finally obtain
\be
G_k(\re^{-\frac{s}{\sigma_k}}) = \sqrt{2} \ (2k)^{k} \ \re^{-k}
\ \re^{\frac{s^2}{2}-\frac{s}{2}\sqrt{5k}}
\ \qua{1+O(\frac{1}{\sqrt{k}})}\ ,
\ee
and thus, for asymptotically large values of $k$
\be
\phi_k(s)=\re^{\frac{s^2}{2}}
\qua{1+O(\frac{1}{\sqrt{k}})}\ .
\ee 
Indeed, by comparison with eq. (\ref{laplace2}), we have thus proven 
that $P(t)$ 
as defined in equation (\ref{asymprob}) is a standard Gaussian 
distribution:
\be
P(t)=\frac{1}{\sqrt{2\pi}} \ \re^{-\frac{t^2}{2}}\ .
\label{finaldistrib}
\ee

This result gives us further information about the asymptotic
behaviour of the central moments $m^{(k)}_s$ 
of distribution $P_k(l)$: indeed,
rescaling each central moment $m^{(k)}_s$ by a factor
$\sigma_k^{-s}\sim k^{-\frac{3}{2}s}$ (correspondingly to the rescaling 
of the discrete variable $l$ to the continuous one $t$), and performing
the $k\goto\infty$ limit, we must find the moments of a standard
(unit variance) Gaussian distribution, namely $0$  or $(s-1)!!$, for odd 
or even values of $s$, respectively. This on one hand confirms
rigorously the estimate
$m^{(k)}_s=\sigma_k^{-s}\cdot(s-1)!!\ [1+O(1/k)]$ ($s$ even) and on the  
other hand matches  the guessed behaviour 
$m^{(k)}_s\sim k^{[\frac{3}{2}s]_{i.p.}}$
($s$ odd).

As for  the area versus length distribution, the result we have
found, eq. (\ref{finaldistrib}),
does not agree with the behaviour proposed in Refs. 
\cite{levy,bellissard}; however  
the role of the $D\goto\infty$ limit, inherent to our treatment, must 
be taken into proper account. In this limit, all loops with more
than one plaquette of the subtended surface lying over the same  plane 
are neglected; on the other hand, on  the two-dimensional lattice
considered in  Refs. \cite{levy,bellissard} there is only one possible 
plane. Moreover, having averaged 
over orientations, we restrict ourselves to  the absolute value of area,
while in \cite{levy,bellissard}, algebraic (oriented) area is 
considered.

When considering very compact loops, their distribution is ruled
by the left tail of eq. (\ref{finaldistrib}). In Ref. \cite{nieuwenhuizen}
the different but somehow related problem of the distribution
$p(s,t)$ of random walks  of $t$ steps visiting exactly $s$ different
sites was addressed on a  cubic lattice. 
In particular, an asymptotic formula   in the regime
of  large $t$ and relatively small $s$ was proposed,  which describes 
the distribution of compact walks.
However the two distributions  are not comparable: indeed, in our 
situation of infinite dimensionality, the most
compact loops of length $2k$ have zero enclosed area, but still visit
a relatively large number of distinct sites, at least $k+1$;
more precisely,  when the dimensionality of the lattice is 
very large, the number
of walks which take more than one step in the same direction
is of order $1/D$ with respect to those stepping in $k$ different 
directions, and therefore even the most compact walk, the one 
coming  back to the origin every two steps, at leading order in $D$
visits $k$ different sites, in addition to the origin. 
In other words, the connection between the two notions of compactness,
as small enclosed area, or as small number of visited sites,
is lost in the limit of infinite dimensionality.

\section{An Infinite Hierarchy of Nontrivial Identities}

We would like to conclude with a simple consideration,  which in our 
opinion is however rather striking and deserve further investigation.
The exact evaluation of moments $M_s^{(k)}$ gives as a byproduct
an infinite set of nontrivial and somehow intriguing identities, the 
simplest of them beeing:
\be
\sum_{l=0}^{k(k-1)/2} G_{k,l} = (2k-1)!! \spzg k=0,1,2,\dots
\ee
and in general
\be
\sum_{l=0}^{k(k-1)/2} G_{k,l} \ l^s = M_s^{(k)} \ (2k-1)!! \spzg k=0,1,2,\dots
\ee
with  $ G_{k,l}$ and  $M_s^{(k)}$ explicitely given by formulae
(\ref{gkl}) and (\ref{explicitmoment1}-\ref{explicitmoment2}), respectively.
These identities are of course encoded in the Jacobi Theta function,
essentially in its modular invariance properties, eq. (\ref{modinv}),  
but they come nevertheless rather unexpected.

Other sets of identities can be obtained if we are somehow able to compute
the $G_{k,l}$ in a indipendent  and simple way. This is indeed very 
easy e.g.
for $G_{k,k(k-1)/2}$, which counts the number of diagrams with $2k$
external points and maximal crossing number, and is thus  obviously
equal to $1$ for any $k$ (just connect the $j^{th}$ point with the 
$(j+k)^{th}$ one, $j=1,\dots,k$). From eq. (\ref{gkl}) we may therefore 
write:
\be
\sum_{i=0}^{k-1}(-1)^i \ton{{k(k+1)/2-1-i(i+1)/2}\atop{k-1}}
\ton{{2k+1}\atop{k-i}}\frac{2i+1}{2k+1}=1  \spzg k=0,1,2,\dots
\label{trivident}
\ee
Identities of this second sort are  however more obvious:
eq. (\ref{trivident}) is just the simplest example of a whole 
bunch of identities
encoded in an obvious simmetry of polynomial $G_k(q)$: 
by applying the Theorem of Section \ref{solution} (eq. (\ref{theorem}))
to  the new polynomial ${\tilde G}_k(q)=q^{k(k-1)/2}G_k(1/q)$,
with $G_k(q)$ expressed in terms of eq. (\ref{gklsum})
one obtains new coefficients ${\tilde G}_{k,l}$,
which obviously satisfy  ${\tilde G}_{k,l}=G_{k,(k(k-1)/2)-l}$;
setting in particular $l=0$, we readily get eq. (\ref{trivident}). 

\section{Conclusions}

The problem of counting closed
paths  of given length  on a lattice, according to the area they enclose,
is a difficult one. In the present paper  we have tackled the question on 
an hypercubic lattice, in the limit of infinite dimensionality. 
Moreover an average over orientations is inherent to the  method
we have used. But these conditions have allowed us to give a closed 
answer, which is moreover exact under the restriction we have assumed.
The main flaw of the whole method resides in our opinion into the fact 
that the computation of even only the first corrections  in $1/D$ 
to coefficients $G_{k,l}$ is out of reach.
On the other hand,  we have presented an explicit solution 
to a combinatorial problem which could hardly be addressed directly;
in this respect, the connection between the abstract combinatorial puzzle
and the physical model proposed in \cite{parisi1} is crucial.
The exact solution proposed for coefficients $G_{k,l}$ has been 
investigated in detail; their asymptotic behaviour has been evaluated;
finally, as a byproduct, an infinite set of new nontrivial identities 
has been obtained.

\bigskip
\noindent{\large\bf  Acknowledgements}
\medskip

We would like to thank Andrea Cappelli and Stefano Ruffo
for fruitful discussions, Vladimir Rittenberg for suggesting us the
use of the Laplace Transform method (Section 6),
and Giuseppe Santoro for informing us of Ref. 
\cite{pauling} and its connection with our work. 

\bigskip
\noindent{\Large\bf  Appendix A}
\medskip

We want to build up a procedure to compute
in a systematic way derivatives of  arbitrary order of $G_k(q)$ with 
respect to $\log q$, at $q=1$. From the explicit expression (\ref{gkq}) of 
$G_k(q)$, we have to take into account both the $q$ dependence
of the integration interval and of the integration 
measure $\nu_q(x)$, given by eq. (\ref{qmeasure}). 
As for the latter,  the task is indeed non completely straightforward, since 
for fixed values of $x$,  $\nu_q(x)$ develops an essential singularity
in the limit $q\goto 1$,
and should somehow be regularized. In presence of an essential singularity,
the result of the limiting procedure depends from the direction chosen
to approach the singularity, and a  regularization is a prescription
on how the singularity is approached. Indeed, if 
 we restrict $q$ to the real axis and approach  $q=1$ 
from smaller values,  it is possible to write the 
integration measure
 $\nu_q(x)$  ($x=\frac{2}{\sqrt{1- q}}\cos\theta$) as a sum of a completely 
regular contribution,  Taylor expandable in the real variable $\eps$, 
($q=\re^{-\eps}$)
plus a singular contribution vanishing with all its derivatives for 
$\eps\goto 0^+$. Indeed we shall show that:
\be
\nu_q(x)= {\cal F}(x,\eps) \ \qua{1+ O(\re^{-1/\eps})}\ ,
\label{asymptoticform}
\ee
 ${\cal F}(x,\eps)$ being an analytic function of $\eps$.

Using modular invariance, we can write
\be
\Theta_1\ton{\frac{\theta}{\pi}|\tau}=\frac{i}{\sqrt{-i\tau}} 
\ \re^{-i\theta^2/\pi\tau}\ 
\Theta_1\ton{\frac{\theta}{\pi\tau}|-\frac{1}{\tau}}\ ;
\label{modinv}
\ee
on the other hand, from definition (\ref{theta}), letting
$\tau=\frac{i \eps}{2\pi}$,
\be
\Theta_1\ton{\frac{\theta}{\pi\tau}|-\frac{1}{\tau}}=
\frac{1}{i}\ \re^{-\frac{\pi^2}{2\eps}}\ \re^{\frac{2\pi\theta}{\eps}}
\ \sum_{n\in\Z} (-1)^n \ 
\re^{-\frac{2\pi^2}{\eps}(n^2+n -2n\frac{\theta}{\pi})}\ .
\ee
For arbitrary small positive $\eps$, and for $\frac{\theta}{\pi}$ 
in the open interval
$(0,1)$, the leading contribution is given by the $n=0$ term in the 
infinite sum, all other terms being exponentially small. We may therefore
write, for $\tau=\frac{i \eps}{2\pi}$, and $0< \eps\ll 1$:
\be
\Theta_1\ton{\frac{\theta}{\pi\tau}|-\frac{1}{\tau}}\sim
\frac{1}{i}\ \re^{-\frac{\pi^2}{2\eps}}\ \re^{\frac{2\pi\theta}{\eps}}
\ \qua{1+ O(\re^{-\frac{1}{\eps}})} \ ,
\ee
and for the measure $\nu_q(x)$ we readily get, in the limit $q\goto 1^-$,
or $q=\re^{-\eps}$, $\eps\goto 0^+$, the approximate behaviour 
(\ref{asymptoticform}), with 
\bea
{\cal F}(x,\eps) &=& \sqrt{1-\re^{-\eps}} \ \re^{\frac{\eps}{8}}
\ \frac{1}{\sqrt{2\pi\eps}}\ \exp\qua{-\frac{2}{\eps}
\ton{\theta-\frac{\pi}{2}}^2 }\\
&=&\sqrt{1-\re^{-\eps}} \ \re^{\frac{\eps}{8}}
\ \frac{1}{\sqrt{2\pi\eps}}\ \exp\qua{-\frac{2}{\eps}\arcsin^2
\ton{\frac{x}{2}\sqrt{1-\re^{-\eps}}}} \label{effedixeps}\\
&=& \frac{1}{\sqrt{2\pi}}\ \re^{-\frac{x^2}{2}} \ \sum_{r=0}^{\infty}
\frac{c_r(x)}{r!} \ \eps^{r}\ ,
\eea
where the coefficients $c_r(x)$ are given by the Taylor expansion
in the neighbourhood of the origin $\eps=0$ of the analytical function:
\be
\sqrt{1-\re^{-\eps}} \ \re^{\frac{\eps}{8}}\ 
\frac{1}{\sqrt{\eps}}\ \re^{-\frac{1}{2}\qua{\frac{4}{\eps}\arcsin^2
\ton{\frac{x}{2}\sqrt{1-\re^{-\eps}}}-x^2}}\ ;
\label{functione}
\ee
$c_0(x)=1$ trivially while 
 higher coefficients $c_r(x)$  can be shown to be 
even polynomials of degree $4r$  in $x$.
In the neighbourhood of $q=1$ we may therefore write, up to terms
vanishing exponentially for small $\eps$:
\be
G_k(q)=\frac{1}{\sqrt{2\pi}}
\int_{-\frac{2}{\sqrt{1-q}}}^{\frac{2}{\sqrt{1-q}}} 
\ dx \ x^{2k}\  \re^{-\frac{x^2}{2}} \sum_{r=0}^{\infty}
\frac{c_r(x)}{r!} \ \eps^{r}\ .
\ee
Let us now observe that the  integration limits diverge as 
$1/\sqrt{\eps}$, while the integrand falls down as a quadratic exponential.
Therefore, up to terms  which vanish exponentially for small $\eps$,
the integration interval can readily be extended to the whole
real axis, and the computation of moments (\ref{momenta}) is finally
reduced to the Taylor expansion of (\ref{functione}) and the evaluation 
of a sum of standard Gaussian integrals:
\be
M_s^{(k)}=\frac{(-1)^s}{(2k-1)!!}
\ \frac{1}{\sqrt{2\pi}}\ \int_{-\infty}^{+\infty} dx 
\ \re^{-\frac{x^2}{2}}\ x^{2k} \ c_s(x)\ .
\ee

As an example, let  us compute explicitly the first moment, $M_1^{(k)}$.
The first coefficient of the Taylor expansion of (\ref{functione})
is 
\be
c_1(x)=-\frac{1}{8} +\frac{1}{4}x^2-\frac{1}{24}x^4\ ,
\ee
and therefore, denoting with $\la\la\dots\ra\ra$ the Gaussian average (with 
variance equal to $1$),
and exploiting the standard relation $\la\la x^{2k}\ra\ra = (2k-1)!!$
\bea
M_1{(k)} &=&\frac{1}{\la\la x^{2k}\ra\ra}\qua{\frac{1}{8}\la\la x^{2k}\ra\ra
-\frac{1}{4}\la\la x^{2k+2}\ra\ra +\frac{1}{24}\la\la x^{2k+4}\ra\ra}\\
&=&\frac{1}{8}-\frac{1}{4}(2k+1)+\frac{1}{24}(2k+3)(2k+1)\\
&=& \frac{k(k-1)}{6} \ .
\eea

\end{document}